\documentclass[superscriptaddress,twocolumn,aps,prl,showpacs]{revtex4}
\usepackage{epsfig}
\usepackage{graphicx}
\usepackage{dcolumn}
\usepackage{bm}
\begin{document}

\title{Random walks with preferential relocations to places visited in the past
and their application to biology}

\author{Denis Boyer}
\email{boyer@fisica.unam.mx}
\affiliation{Instituto de F\'\i sica, Universidad Nacional Aut\'onoma de 
M\'exico, D.F. 04510, M\'exico}
\affiliation{Centro de Ciencias de la Complejidad, Universidad Nacional 
Aut\'onoma de M\'exico, D.F. 04510, M\'exico}

\author{Citlali Solis-Salas}
\affiliation{Instituto de F\'\i sica, Universidad Nacional Aut\'onoma de 
M\'exico, D.F. 04510, M\'exico}

\date{\today}

\begin{abstract}
Strongly non-Markovian random walks offer a promising modeling framework for 
understanding animal and human mobility, yet, few analytical results
are available for these processes. Here we solve exactly a model with 
long range memory where a random 
walker intermittently revisits previously visited sites according to a 
reinforced rule. The emergence of frequently visited locations generates 
very slow diffusion, logarithmic in time, whereas the walker probability 
density tends to a Gaussian. This scaling form does not emerge from 
the Central Limit Theorem but from an unusual balance between random and 
long-range memory steps. In single trajectories, occupation patterns are 
heterogeneous and have a scale-free structure. The model exhibits good 
agreement with data of free-ranging capuchin monkeys.
\end{abstract}

 \maketitle

The individual displacements of living organisms exhibit rich statistical 
features over multiple temporal and spatial scales.  
Due to their seemingly erratic nature, animal movements 
are often interpreted as random search processes
and modeled as random walks \cite{turchin,colding,randomsearch}.
In recent years, the increasing availability of data on animal 
\cite{interm4,nathan,sims,weimer} as well as human 
\cite{geisel,gonzalez,song1,song} 
mobility have motivated numerous models
inspired from the simple random walk (RW). Let us mention, in particular,
multiple scales RWs, such as L\'evy walks \cite{levy1,levy2}
or intermittent RWs \cite{interm4,interm1,interm2,interm3}, which are 
walks with short local movements mixed with less frequent but
longer commuting displacements.

Markovian RWs are the basic paradigm for modeling animal and 
human mobility and they provide useful insights
at short temporal scales. However, empirical studies conducted 
over long periods of times reveal 
pronounced non-Markovian effects \cite{song,chinos,interface}. 
As for humans, mounting evidence shows that many
animals have sophisticated cognitive abilities and use memory to move 
to familiar places that are not in their immediate perception range 
\cite{fagan,janson}. The use of long-term memory should
strongly impact movement and it is probably 
at the origin of many observations which
are incompatible with RWs predictions, such as, very slow diffusion, 
heterogeneous space use, the tendency to revisit often particular places 
at the expense of others, or the emergence of 
routines \cite{gautestad2005,gautestad2006,song1,song,chinos,interface}. 
Non-Markovian random walks where movement steps depend on the whole
path of the walker \cite{elephant1,elephant2,italian}
offer a promising modeling framework in this context. But the relative
lack of available analytical results in this area limits the understanding
of the effects of memory on mobility patterns.

Self-attracting or reinforced RWs are an important class of non-Markovian 
dynamics \cite{annals}. In these processes, typically, a walker on a 
lattice moves to a nearest-neighbor
site with a probability that depends on the number of times this site  
has been visited in the past \cite{davis,siam,epl}. These walks must be in 
principle described by a hierarchy of multiple-time distribution functions, 
or can be studied within field theory approaches \cite{nuovo}. 
In a slightly different context, 
some exact results have been
obtained for the mean square displacement (MSD) in globally reinforced
models, such as the so-called elephant walk 
\cite{elephant1,elephant2}, where the walker tends to move
in the same direction than the sum of all its previous movement steps. 

\begin{figure}
  \centerline{\includegraphics*[width=0.35\textwidth]{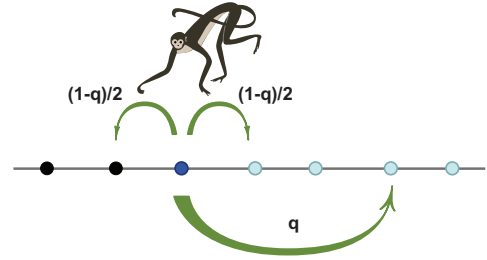}}
  \caption{(Color online) A model walker combining random steps to nearest
  neighbor sites and relocations, at a rate $q$, to sites 
  visited in the past (marked in light color).}
\label{fig1}
\end{figure}
In this Letter we solve a minimal, lattice version of a reinforced model 
proposed some 
time ago in the ecological literature \cite{gautestad2005,gautestad2006}, 
where a walker can either move randomly (explore locally) or stochastically
relocate to places visited in the past (via long distance steps). 
A constant parameter describes the relative rate of these two movement 
modes (Fig. \ref{fig1}). This RW model with long range memory is,
to our knowledge, one of the very 
few where not only the MSD is derived exactly, but also
the asymptotic form of the full probability density. We then compare the model 
with field data and infer the strength of memory use in real animals.

We consider a walker with position ${\mathbf X}_t$ at time $t$ on a regular 
$d$-dimensional lattice with unit spacing, and initially located at 
${\mathbf X}_0={\mathbf 0}$. 
Consider $q$ a constant parameter, $0<q<1$. 
At each discrete time step, $t\rightarrow t+1$, the walker decides with 
probability $1-q$ to visit 
a randomly chosen nearest neighbor site, as in the standard RW. 
With the complementary probability $q$, the walker relocates directly 
to a site visited in the past (Fig. 1). In this case,
the probability to choose a given lattice site, among all the visited sites, 
is proportional to the number 
of visits this site has already received in the interval $[0,t]$. 
It is thus more likely to revisit a site which has been 
visited many times than a site visited only once. This linear preferential 
revisit rule is
equivalent to choosing a random integer $t^{\prime}$ uniformly in the interval 
$[0,t]$ and to return to the site occupied at $t^{\prime}$. This model bears 
some similarities with that of ref. \cite{evansmaj}, where a RW is 
stochastically \lq\lq reset" to the origin ($t'=0$) at a constant rate. 
Here, the RW can be reset to any previous time, or visited site, thus making 
the process highly non-Markovian. 

We summarize our main results in $1d$ for this model, where
memory profoundly modifies the normal diffusion process and generates
complex patterns of space occupation. (The results naturally extend to 
higher dimensions.) Let $P(n,t)$ be the probability that $X_t=n$. The MSD, 
defined as the ensemble average
$\langle X_t^2\rangle=M_2(t)\equiv\sum_{n=-\infty}^{\infty}n^2P(n,t)$,
is calculated exactly for all $t$. Asymptotically, it grows very slowly
with time:
\begin{equation}\label{m2tas}
M_2(t)\simeq\frac{1-q}{q}\left[\ln (qt)+\gamma \right],\quad t\gg1,
\end{equation}
with $\gamma=0.5772...$ the Euler constant. In addition,
the distribution $P(n,t)$ tends to a 
Gaussian, as in normal diffusion, but with a variance given by the
anomalous logarithmic law (\ref{m2tas}) instead of the
usual normal law $\propto t$:
\begin{equation}\label{gauss}
P(n,t)\rightarrow G(n,t)\equiv\frac{1}{\sqrt{2\pi M_2(t)}}\ 
e^{-\frac{n^2}{2M_2(t)}},
\end{equation}
Notably, the mechanism that makes the process eventually Gaussian is driven 
by memory and thus markedly differs from the Central Limit 
Theorem. In particular, the convergence toward this scaling form is 
logarithmically slow in time, thus it is difficult 
to observe in practice in discrete time simulations. 
%
%
We also study the probability $P^{(v)}_t(m)$ 
that a site, randomly chosen among the sites visited by a single walker
in $[0,t]$, has received exactly $m$ visits.
Numerical results in $2d$ suggest a power-law behavior:
\begin{equation}\label{visitspdf}
P^{(v)}_t(m)\propto m^{-\alpha},\quad {\rm with}\quad\alpha\simeq 1.1,
\end{equation}
which indicates that the walker occupies space in a very 
heterogeneous way. The model in $2d$ agrees quantitatively with 
trajectory data of capuchin monkeys ({\it Cebus capucinus}) in the wild.

We next present a derivation of the results. In contrast with most
path-dependent processes, here, a closed and exact master equation can 
be written for $P(n,t)$, see the Supplemental Material. In $1d$, it reads:
\begin{eqnarray}\label{logreset}
P(n,t+1)&=&\frac{1-q}{2}P(n-1,t)+\frac{1-q}{2}P(n+1,t)\nonumber\\
&+&\frac{q}{t+1}\sum_{t'=0}^tP(n,t').\
\end{eqnarray}
The last term in Eq.(\ref{logreset}) indicates that site $n$ can
be visited (from any other site) following the memory rule provided that
the walker was at $n$ at an earlier time $t'$.

We define the even moments of the distribution as
$M_{2p}(t)=\sum_{n=-\infty}^{\infty} n^{2p}P(n,t)$
with $p$ a positive integer ($M_{2p+1}(t)=0$ by symmetry). 

\begin{figure}
  \centerline{\includegraphics*[width=0.48\textwidth]{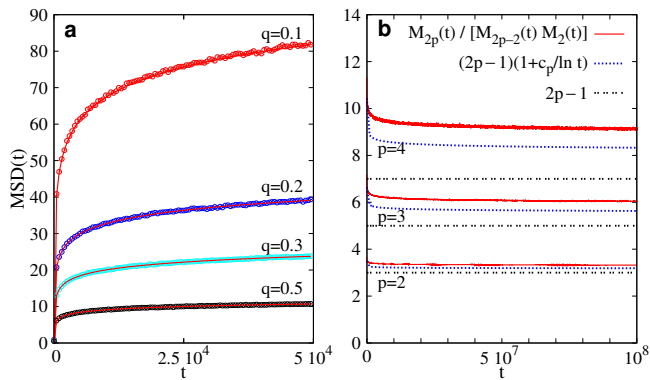}}
  \caption{(Color online) a) MSD as a function of time 
  for different memory strengths $q$. Symbols represent simulation 
  results and the solid lines Eq. (\ref{m2tas}). b) Time evolution 
  of the moment ratio for $p=2,3$ and $4$ from
  simulations with $q=0.1$ (solid red line) and the first order calculation 
  Eq. (\ref{quotient}) (dotted line). The curves tend to
  $2p-1$ at very large $t$.}
\label{fig2}
\end{figure}

{\it Mean square displacement.$-$}Taking the second moment 
of Eq.(\ref{logreset}), we obtain an evolution equation for the MSD:
\begin{equation}\label{m2_1}
M_2(t+1)=1-q+(1-q)M_2(t)+\frac{q}{t+1}\sum_{t'=0}^{t}M_2(t'),
\end{equation}
where we have used the normalization condition $M_0(t)=1$. 
The above equation can be solved by introducing the Z-transform 
of $M_2(t)$, defined as $\widetilde{M}_2(\lambda)
=\sum_{t=0}^{\infty} \lambda^tM_2(t)$. Transforming Eq.(\ref{m2_1}) and
using the identity 
$\lambda^t/(t+1)=\lambda^{-1}\int_0^{\lambda}u^tdu$, 
one obtains:
\begin{equation}\label{odem2}
\frac{\widetilde{M}_2(\lambda)}{\lambda}=\frac{1-q}{1-\lambda}+(1-q)
\widetilde{M}_2(\lambda) 
+\frac{q}{\lambda}\int_0^{\lambda}du 
\frac{\widetilde{M}_2(u)}{1-u}. 
\end{equation}
This equation becomes a first order ordinary differential equation
after taking a derivative with respect to $\lambda$.
As $M_2(t=0)=0$, the condition to be fulfilled by the
solution of Eq. (\ref{odem2}) is 
$\widetilde{M}_2(\lambda=0)=0$. One finds:
\begin{equation}\label{m2lambda} 
\widetilde{M}_2(\lambda)=-\left(\frac{1-q}{q}\right) 
\frac{\ln(1-\lambda)-\ln[1-(1-q)\lambda]}{1-\lambda}. 
\end{equation} 
The function $f(t)$ such that
$\sum_{t=0}^{\infty}\lambda^t f(t)=\ln[1-(1-q)\lambda]/(1-\lambda)$, 
is $f(t)=-\sum_{k=1}^t (1-q)^k/k$.
Therefore, Eq.(\ref{m2lambda}) can be inverted, giving the exact solution:
\begin{equation}\label{exactM2}
M_2(t)=\frac{1-q}{q}\sum_{k=1}^t\frac{1-(1-q)^k}{k}.
\end{equation}
At large $t$, $\sum_{k=1}^{t}1/k\simeq \ln t+\gamma$ and
$\sum_{k=1}^t(1-q)^k/k\simeq -\ln q$, yielding the
asymptotic behavior (\ref{m2tas}) up to order $(\ln t)^0$. This
result holds in any spatial dimensions.
Figure \ref{fig2}a displays Eq.(\ref{m2tas}) for several values of $q$, 
in very good agreement with numerical simulations. Despite of the random
steps, at any finite $q$, memory induces frequent returns to the same sites 
and very slow diffusion.

{\it Higher moments.$-$}The asymptotic form of the propagator 
$P(n,t)$ can be extracted in 
principle from the knowledge of all its moments at large $t$. We first assume 
that a scaling relation is satisfied for $t$ large enough:
\begin{equation}\label{scalingmem}
M_{2p}(t)\simeq a_{p}[M_2(t)]^{p},
\end{equation}
for any integer $p$, with $a_p$ a constant. 
A well-known property of the Gaussian distribution with zero mean 
and arbitrary variance
is that of having $a_0=1$ and
\begin{equation}\label{recugauss}
a_p=(2p-1)a_{p-1} ,\ \ p\ge 1.
\end{equation}
We take the $2p$-th moment of Eq.(\ref{logreset}): 
\begin{eqnarray}\label{expmommem}
M_{2p}(t+1)&-&M_{2p}(t)=1-q+(1-q)\sum_{k=1}^{p-1}C_{2p}^{2k}M_{2k}(t)\nonumber\\
&+&\frac{q}{t+1}\sum_{t'=0}^{t} [M_{2p}(t')-M_{2p}(t)]. 
\end{eqnarray}
Since $M_2(t)$ diverges at large $t$, from
(\ref{scalingmem})
the leading term in the first sum of (\ref{expmommem}) is that with $k=p-1$, 
like in the simple RW. But unlike in the RW, the left-hand-side
$M_{2p}(t+1)-M_{2p}(t)\rightarrow0$ and can be neglected, since it is 
$\simeq dM_{2p}/dt\propto (\ln t)^{p-1}/t$.
Thus, using Eqs. (\ref{scalingmem}) and (\ref{m2tas}), Eq. (\ref{expmommem})
gives the following relation for the $a_p$'s:
\begin{equation}\label{m2p2mem}
a_p= p(2p-1)a_{p-1} \lim_{t\rightarrow \infty}
\frac{(t+1)(\ln t)^{p-1}}{\sum_{t'=cst}^{t'=t}[(\ln t)^p-(\ln t')^p]}.
\end{equation}
The limit in (\ref{m2p2mem}) turns out to be $1/p$ \cite{abram}. 
Therefore relation (\ref{recugauss}) is obtained, implying
the Gaussian form (\ref{gauss}). This analysis illustrates that, here,
Gaussianity is not the result of 
random increments producing fluctuations that scale as $\sqrt{t}$, but 
rather emerges in a process with very small fluctuations (of order 
$\sqrt{\ln t}$) from a balance between purely random steps
and recurrent memory steps.
 
To examine how $a_p/a_{p-1}$ converges towards $2p-1$, we
relax the condition that $a_p$ is constant. Assuming that 
$da_p(t)/dt$ does not decay slower than an inverse power law of time, 
one can still neglect the left-hand-side of (\ref{expmommem}). Keeping the
terms of order $(\ln t)^{p-1}$ and $(\ln t)^{p-2}$,
the leading time-dependent correction is obtained:
\begin{equation}\label{quotient}
\frac{a_p}{a_{p-1}}(t)=(2p-1)\left(1+\frac{c_p}{\ln t}\right)+O((\ln t)^{-2})
\end{equation}
with $c_p=(p-1)[1+q/6(1-q)]$.
We see from (\ref{quotient}) that the distribution $P(n,t)$ converges 
{\it extremely} slowly toward the Gaussian form (typically after 
$t\sim 10^{100}$), due to corrections of order 
$1/\ln t$ in the moment relations. In standard sums of random 
variables, these corrections are $O(1/\sqrt{t})$.
Figure \ref{fig2}b displays the quotient 
$Q_p(t)\equiv M_{2p}(t)/[M_{2p-2}(t)M_2(t)]$ obtained from 
numerical simulations as a function of time, for $p=2,3$ and $4$.
If a scaling relation (\ref{scalingmem}) strictly holds, $Q_p(t)=a_p/a_{p-1}$. 
At $t=10^8$, $Q_p(t)$ still differs 
significantly from $2p-1$. Fig. \ref{fig2}b also
displays $\frac{a_p}{a_{p-1}}(t)$ as given by formula (\ref{quotient}). 
What seems to be a plateau at a constant value $>2p-1$ is actually 
a very slowly decaying function. The differences between the simulation and
the analytical results are due to terms $O((\ln t)^{-2})$ or higher, which
are not that small.

\begin{figure}
  \centerline{\includegraphics*[width=0.5\textwidth]{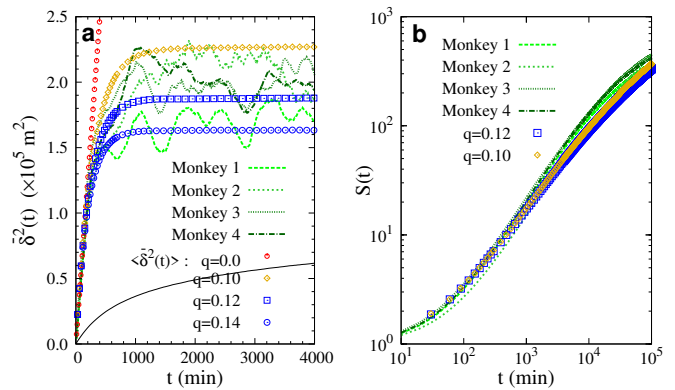}}
  \caption{(Color online) a) TASD of radio-collared capuchin monkeys 
  (dotted lines) and of the model walker (symbols) in $2d$ with the time 
  step set to $\Delta t=30$ min and the cell size to $50$m$\times50$m. The
  solid black line is the MSD of the model with $q=0.12$. b) Number of 
  distinct cells visited by the monkeys and the simulated
  model walker with the parameters of a).}
\label{fig3}
\end{figure}
{\it Monkey mobility data.$-$}The very slow growth of the MSD with $t$ 
in our model agrees qualitatively with the fact that most animals have
limited diffusion or home ranges
\cite{moorcroftlewis,gautestad2005,borger,vanmoorter,interface}. 
We further compare the model predictions with
trajectories of real animals in the wild.
The displacements of four radio-collared capuchin monkeys were recorded 
during a period of six months in Barro Colorado Island, Panama. 
Discretized $2d$ positions,
with resolution $\Delta \ell=50$ m were recorded every 10 min 
(see \cite{megpnas,interface} for details). Since no ensemble averages
can be performed, we calculated for each individual monkey
the time-averaged square displacement (TASD), noted as
$\overline{\delta^2}(t)$, along each trajectory \cite{interface}. 
We also calculated this quantity
for simulated $2d$ walks in the model:
\begin{equation}\label{tasd}
\overline{\delta^2}(t)\equiv \frac{1}{N-t}\sum_{i=1}^{N-t}|{\mathbf X}_{i+t}
-{\mathbf X}_{i}|^2,
\end{equation}
with $N$ the total number of positions, and then
obtained the numerical $\langle \overline{\delta^2}(t)\rangle$
by averaging over many walks. This quantity is {\it a priori}
different from the MSD.

In Fig. \ref{fig3}a, the animals have a Brownian
regime with $\overline{\delta^2}(t)\simeq4Dt$
at short times, with a diffusion coefficient $D\simeq 300$ m$^2/$min
for all four monkeys, followed by a saturation at a roughly constant value. 
Setting the lattice spacing $\Delta \ell=50$ m in the model, too,
the model time step is adjusted to $\Delta t=30$ min so that
$\langle\overline{\delta^2}(t)\rangle$ matches the monkeys TASD at short times
(hence, the 6 months of foraging data correspond to
$N=8640$). 
These parameters being fixed, the value $q\simeq 0.12\pm 0.02$ best describes 
the monkeys TASD over the entire time range (Fig. \ref{fig3}a). 
The resulting relocation rate 
$r\equiv q/\Delta t\simeq 0.004$ min$^{-1}$ is low, suggesting that
memory use by capuchin monkeys is intermittent. 
But even this small $r$ 
strongly affects diffusion after a few hours.

We note at this point that the model is non-ergodic, in the sense that
$\langle \overline{\delta^2}(t)\rangle\ne M_2(t)$ \cite{godec}. Here
$\langle \overline{\delta^2}(t)\rangle$ quickly reaches a plateau, 
whereas $M_2(t)$ is smaller and slowly grows with $t$, 
as shown in Fig. \ref{fig3}a with $q=0.12$. Another criteria of
non-ergodicity involves an ergodicity breaking
parameter, which measures the fluctuations among time averages 
obtained from different trajectories: $EB\equiv\langle
[\overline{\delta^2}(t)]^2\rangle/\langle\overline{\delta^2}(t)\rangle^2-1$
\cite{he,barkai}.
According to this criteria, a process is non-ergodic if 
$\lim_{N\rightarrow\infty}EB\ne0$. Setting $q=0.12$ and 
$N=8640$ gives $EB=0.055$ for the model. We actually find 
that $EB\rightarrow0$ as $N\rightarrow\infty$ (not shown). Hence, the model is
ergodic in this second sense: different long trajectories
have the same $\overline{\delta^2}(t)$ at short times (or 
$\overline{\delta^2}(t)\simeq\langle\overline{\delta^2}(t)\rangle$). 
Interestingly,
this similitude is also observed in the four monkeys (see Fig. \ref{fig3}a).

\begin{figure}
  \centerline{\includegraphics*[width=0.4\textwidth]{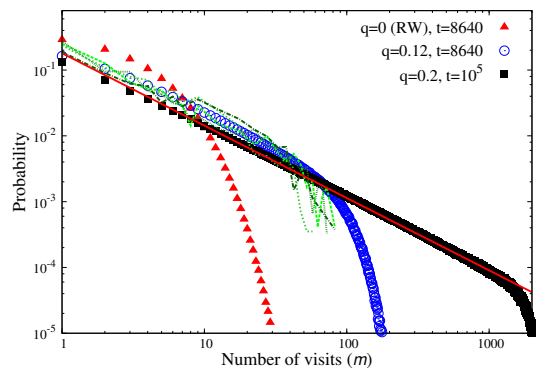}}
  \caption{(Color online) Probability that a visited site chosen at random
  has been visited $m$ times in $t$ walker steps. Open circles: model
  simulations with $q$ fitted in Fig. \ref{fig3}a and $t$
  corresponding to 6 months. The monkey data is shown in 
  dashed green lines. The solid (red) line has slope $-1.10$.}
\label{fig4}
\end{figure}
Figure \ref{fig3}b shows the number $S(t)$ of distinct sites visited by the
model walker in $2d$ with the parameters fitted above, confirming the good 
agreement with empirical data. 
Further insight into the recurrent properties of these walks is given
by the distribution function of the number $m$ of visits per site,
$P^{(v)}_t(m)$. Scale-free distributions often are an outcome of 
preferential rules, such as in Yule processes \cite{yule,simon} or network 
growth models with preferential attachment \cite{barabasi,leyvraz}. 
In a model trajectory, many
sites are visited only once whereas fewer sites are visited very often
and thus likely to be visited again, giving rise to the
formation of \lq\lq hot-spots" of activity.
We speculate that $P^{(v)}_t(m)$ in $2d$ is scale-free when $q\ne0$.
The exponent $\alpha$ introduced in Eq.(\ref{visitspdf}) 
seems to be independent of $q$, as shown in Figure \ref{fig4}. 
The scaling regime is more extended for $q$ and $t$ large.
Monkeys visitation patterns closely follow the theoretical law. 
Unlike in Yule processes or the reinforced walk with preferential
visits of ref. \cite{song}, 
spatial correlations are strong here (the sites near a hot-spot are likely 
to be visited often, too), making the analytical calculation
of $P^{(v)}_t(m)$ quite challenging.

{\it Discussion.$-$}Motivated by the modeling of animal mobility, 
we have studied a minimal, solvable random walk model with infinite 
memory where the sites visited in the past are preferentially revisited. 
Memory induces very slow diffusion and slowly drives the process towards
Gaussianity. This latter form contrasts with the scaling functions
of Markovian RW models exhibiting logarithmic diffusion ({\it e.g.}, the Sinai 
model \cite{sinai1,sinai2,sinai3}) or stopped diffusion ({\it e.g.}, 
the RW stochastically reset to the origin \cite{evansmaj}), which 
have exponential tails. Likewise, the scaling function
of the elephant walk model \cite{elephant1} in the anomalous regime is not 
Gaussian, although its precise form is not known \cite{gandhiPRE}.
Our results point out a new mechanism for the 
emergence of Gaussian distributions, which could be generic in stochastic 
processes where a recurrent memory does not prevent fluctuations from
diverging with time, but make them grow slower than a power-law. As a 
consequence, the process is asymptotically 
described by an effective Fokker-Planck equation with a time dependent 
diffusion coefficient, $D=\frac{1-q}{2qt}$, see Eqs. (\ref{m2tas})-(\ref{gauss}). 
Such an effective description is useful for studying
first-passage properties \cite{effecFP}. The aging properties of the model
also deserve further study.

The primate mobility data presented here provide additional
evidence that memory is a key factor for home range 
self-organization \cite{gautestad2005,gautestad2006,borger,
vanmoorter,boyerwalsh,oikos2013}. Our model
suggests that the use of memory
is likely to be intermittent in animals, and that even a very
small rate $r$ can induce very slow diffusion and heterogeneous
patterns of space occupation.

We thank  M.C. Crofoot, L. Lacasa, H. Larralde, F. Leyvraz, G. Oshanin, 
I. P\'erez-Castillo, A. Robledo, S. Thurner and P.D. Walsh for many
fruitful discussions and D. Aguilar for technical support. 
This work was supported by 
the Grant IN103911 of the Universidad Nacional Aut\'onoma de M\'exico.


\end{document}